\documentclass[preprint,12pt]{elsarticle}
\usepackage{natbib}
\usepackage{amssymb}
\usepackage{float}
\usepackage{booktabs}
\usepackage{lscape}
\usepackage{rotating}
\usepackage{caption}
\usepackage{subcaption}
\usepackage{longtable}
\usepackage{graphicx}
\usepackage{hyperref}
\journal{Elsevier}
\begin{document}

\begin{frontmatter}



\title{THE INFORMATION FLOW AMONG GREEN BONDS EXCHANGE TRADED FUNDS}


\author[label1]{Wenderson Gomes Barbosa}
\author[label1,label2]{Kerolly Kedma Felix do Nascimento}
\author[label3]{Fábio Sandro dos Santos}
\author[label1]{Tiago A. E. Ferreira}

\affiliation[label1]{organization={UFRPE},%Department and Organization
            addressline={Rua Dom Manoel de Medeiros}, 
            city={ Recife},
            postcode={52171-900}, 
            state={PE},
            country={Brazil}}
\affiliation[label2]{organization={URCA},%Department and Organization
            addressline={Av. Leão Sampaio, 107}, 
            city={Juazeiro do Norte},
            postcode={63041-235}, 
            state={CE},
            country={Brazil}}
\affiliation[label3]{organization={UFPI},%Department and Organization
            addressline={BR-135, KM 3}, 
            city={Bom Jesus},
            postcode={64900-000}, 
            state={PI},
            country={Brazil}}  

\begin{abstract}
  %In this article, we investigate the flow of information between ETFs (Exchange Traded Funds) of Green Bonds from three world markets: the US, the Canadian, and the European markets between 2021 and 2022.
  %%markets, the US market, the Canadian market, and the European market between the years 2021 and 2022. 
  %It was perceived the dominance of the American market among the other two markets, the Canadian and the European market, I tried the FLMB ETF as the largest transmitter of information among the analyzed ETFs, in the Canadian market the HGGB ETF has been highlighted as a major transmitter of information to Europe, but is heavily influenced by the FLMB, the European market the ETFs FLRG and GRON.MI has great prominence in sending information to other ETFs in Europe. KLMH. F in Europe is highlighted as the receiver of information.  Through this article, it was possible to understand the flow of information between the ETF Greenbonds markets and their dimensionality.
  This article investigates the information flow between 13 Green Bond ETFs (Exchange Traded Funds) from three global markets: the USA, Canada, and Europe, between 2021 and 2022. We used the transfer entropy and effective transfer entropy methods to model and investigate the Green Bond price information flow between these global markets. The American market demonstrated market dominance among the other two markets (Canadian and European). The FLMB Green Bond of the American ETF presented the greatest flow of information transfer among the ETFs analyzed, being considered the dominant ETF among the three Green Bond ETF markets investigated. The HGGB ETF has emerged as a major information transmitter in Europe and in the Canadian market, but it has had a strong influence from the American ETF FLMB. In the European market, the FLRG and GRON.MI bonds played a major role in the flow of information sent to other ETFs in Europe. The KLMH.F in Europe is highlighted as the largest receiver of information. Thus, through this article, it was possible to understand the direction of the flow of information between the Green Bond ETF markets and their dimensionality.
\end{abstract}            



%\item We compare results from global markets: the US, Canada, and Europe.
%\item A large flow of information was transferred to Europe from the other two markets.
%\item USA has the dominant ETF among those analyzed.


\end{frontmatter}

\section{INTRODUCTION}\label{sec:intro}
%There is growing concern of political and financial authorities with the problems caused by climate change, as they can cause serious risks to the financial system as a whole, both physical, by the destruction of the infrastructure of cities when financial, because their profit margins can be affected by its consequences \cite{sartzetakis2021green}. With the realization of risks, plans were drawn up to mitigate these effects, it was that Green Bonds appeared, which are fixed-income bonds that have received this name because the issuing institution is committed to use these resources in projects that have a sustainable bias and help mitigate climate change, this is the main difference from traditional bonds, because they do not have this investment focus \cite{bedendo2023greening}.
There is growing concern among political and financial authorities about the problems caused by climate change. Climate change can pose serious risks to the economic system as a whole, both in physical and financial structures, because profit margins can be affected by its consequences \cite{sartzetakis2021green}. With the identification of risks, plans were drawn up to mitigate these effects, thus giving rise to “Green Bonds”. Green Bonds are fixed-income securities that received this name because the issuing institution undertakes to use these resources in projects with a sustainable bias and help mitigate climate change. This is the main difference from traditional bonds, as they do not have this investment focus \cite{bedendo2023greening}.

The first institution to issue Green Bonds was the European Investment Bank in 2007 \cite{antoniuk2021climate} to begin %start 
financing sustainability projects. Later, several multilateral institutions also started issuing, such as the World Bank, the Asian Development Bank, and the African Development Bank. Chambwera \textit{et al.}  \cite{chambwera2014economics} showed that mitigating climate problems will take \$70-100 billion. %show that to mitigate climate problems, it will take \$70-100 billion.

%Large banks and asset managers create ETFs (Exchange Traded Funds), which are investment funds that are created to invest in a portfolio of assets that seek to replicate the portfolio and profitability of a given benchmark index, such as the SP500 \cite{wurgler2010economic}\cite{brown2021etf}, this work aims to study ETFs that in their composition have Green Bonds, so that which person and any level of income had access to invest in this asset that has great impact for the future of the planet and the next generations. 
Large banks and asset managers create ETFs (Exchange Traded Funds), which are investment funds designed to invest in a portfolio of assets to replicate the portfolio and profitability of a given benchmark index, such as the SP500 \cite{wurgler2010economic,brown2021etf}. In this work, we study ETFs with Green Bonds composition. %This work aims to study ETFs that have Green Bonds in their composition, which 
Anyone with any income level can invest in this asset that greatly impacts the future of the planet and the next generations.

Asia has been excelling in the issuance and analysis of Green Bonds. Li \textit{et al.} \cite{li2024impact} showed that companies that issue these bonds in cities in China have increased productivity and reduced debt, obtained cheaper financing, and reduced regulatory and image risks. Taghizadeh-Hesary \textit{et al.} \cite{taghizadeh2021utilising} noted that the Asian Green Bond market has the general characteristic of having high returns coupled with higher risk and greater variability. The study of Abhilash \textit{et al.} \cite{abhilash2023bond} demonstrated that Green Bonds with better ratings have a significant effect on yield. Furthermore, investments made by these securities are more stable and less risky in the long term, so they accept lower profits in exchange for more security. 
 
Tang and Zhang \cite{tang2020shareholders} conducted a study that analyzed the shares of companies from 28 countries that issued Green Bonds between 2007 and 2017. They concluded that the issuance of these securities had a positive impact of $1.4\%$ on the share's price during the period analyzed. However, there was no evidence of greater profitability compared to conventional securities, but rather an increase in the liquidity of shares.

Flammer \cite{flammer2021corporate} reinforced the results of Tang and Zhang \cite{tang2020shareholders}, and the market responded positively to the issuance of Green bonds. It indicates that companies apply resources to improve their environmental indicators and that after the issue, companies can attract more qualified investors with an environmentally conscious profile, even without having a differentiation in the cost of capital, its main contribution being the signaling of environmental improvement.

Daubanes \textit{et al.} \cite{daubanes2021firms} supported the results of Flammer \cite{flammer2021corporate} that Green Bonds signal to markets that the company commits to projects aimed at improving environmental indicators. They also corroborated with Tang and Zhang \cite{tang2020shareholders} that Green Bonds positively impact market shares, especially in sectors that have a high impact on oil derivatives pricing policies, which cause impacts on carbon emissions.
The study conducted by Pham \cite{pham2016risky} makes a comparison between conventional bonds and Green Bonds using multivariate GARCH models and data found in the S\&P Green Bond Index, which found a volatility cluster effect, showing that Green Bonds are susceptible to shocks faced by conventional securities, with differences in magnitude and frequency over time. If using a structural autoregressive vector model (VAR), Reboredo and Ugolini  \cite{reboredo2020price} confirm Pham's theory \cite{pham2016risky} that Green Bonds are susceptible to conventional bond shocks. In their study, the authors compared Green Bonds with American treasury bills, and the exchange market received a lot of influence from these, but with little ability to receive information back. They show a strong influence of the treasury bonds and the foreign exchange market on Green Bonds.

There is great interest in understanding the flow of information in financial and economic systems. He and Shang \cite{he2017comparison} used the transfer entropy method to quantify information transfer between financial time series of nine indices of the USA, Europe, and China stock markets, showing the USA as a major transmitter of information between markets. Jale \textit{et al.} \cite{jale2019information} used the transfer entropy to measure the information flow between the Brazilian market index (Ibovespa) and traded stocks. Caglar and Hancock \cite{caglar2019network} also used the transfer entropy method and the divergence instead of deviation to analyze financial time series. Caserini and Pagnottoni \cite{caserini2022effective} examined the dynamics of information flow between the CDS market (Credit Default Swaps) and sovereign securities of several European countries. During financial crises, the securities market was found to be more efficient in measuring credit risk than the CDS, finding a great indicator of sovereign credit risk, especially in times of crisis. Yijun \textit{et al.} \cite{yijun2023impact} also used information transfer to quantify crisis risks. Their study examined the effect of the COVID-19 crisis on the banking market. During the pandemic, there has been an increase in the complexity of financial operations, with increased vulnerability to external shocks, suggesting a greater need for banking regulation to mitigate the risks caused by increased interaction. Kayal and Maiti \cite{kayal2023examining}, starting from the same principle as Yijun \textit{et al.} \cite{yijun2023impact} to examine the effect of crises on the flow of information, analyzed the gold, silver, and oil markets between the 2008 crises and the COVID-19 crisis, and the strong influence of crises on the volatility of daily returns of these assets.

Thus, it is possible to find many studies on the information flow in financial and economic processes in general. However, when the target is the Green Bonds, it is not easy to find information on flow analysis. The authors thoroughly searched the literature and found nothing about this issue. Therefore, we employ the transfer entropy to analyze the direction of information transfer between Green Bonds ETFs of three markets: the American, Canadian, and European. The transfer entropy method, developed by Schreiber \cite{schreiber2000measuring}, quantifies the flow of information between two time series, which allows us to identify whether the flow is one-dimensional or two-dimensional, its symmetry or asymmetry, and which time series has the power to influence the other series. It is also possible to understand how the information flow between the time series works and analyze the direction of information through the time series.

The paper is organized as follows. Section \ref{sec:metho} presents the definition and description of the data from the research methodology. Section \ref{sec:result} presents the results and discussion. Section \ref{sec:conclu} concludes.

\section{METHODOLOGY}\label{sec:metho}

This work proposes to analyze the information flow of green bonds in the USA, Canada, and European markets. Transfer entropy and effective transfer entropy methods were employed to this end. Those methodologies are described in the following sections. 

\section{Transfer Entropy}

Schreiber \cite{schreiber2000measuring} introduced the Transfer Entropy (TE) theory to measure the flow of directional information between dynamical systems (deterministic and stochastic).
%Introduced the theory of TE by \cite{schreiber2000measuring}, as a measure of directional information flow between dynamical(deterministic and stochastic) systems. 
This method has been used in studies of diverse phenomena such as brain networks \cite{vicente2011transfer}, animal behavior \cite{orange2015transfer} and \cite{butail2014information}, solar wind \cite{wing2016information}, complex networks \cite{lizier2011information}, world wide web dynamics \cite{oka2013exploring} and finances \cite{marschinski2002analysing,kwon2008information,dimpfl2014impact,kwon2008information1,kwon2012asymmetric,jale2019information,nascimento2022has}.%\\

Let two systems be described by observation sequences of length $N$, $X = \{x_t, t = 1,2,...,N \}$ and $Y = \{y_t, t = 1,2,...,N \}$. It is assumed that such systems can be approximated by the stationary Markov process of order $k$ and $l$, respectively, for the sequences $X$ and $Y$. The conditional probability of the state $x_{t+1}$ at the instant $t+1$ is independent of the state $x_{t-k+1}$ as
$$x_{t-k}:p(x_{t+1}\vert x_t,...,x_{t-k+1}) = p(x_{t+1}\vert x_t,...,x_{t-k}),$$ which using $$x^{(k)}_{t} = (x_t,...,x_{t-k+1}),$$ can be written as $$
p\Big(  x_{t+1}\Big\vert x^{(k)}_t\Big) = p\Big(  x_{t+1}\Big\vert x^{(k-1)}_t\Big).$$ Likewise for the system $Y$ we have, $$p\Big(  y_{t+1}\Big\vert y^{(l)}_t\Big) = p\Big(  y_{l+1}\Big\vert y^{(l-1)}_t\Big),$$ where $$y^{(l)}_t = (y_t,...,y_{t-l+1}).$$ 

The TE method takes into account temporal dependencies by incorporating past observations $x^{(k)}_t$ and $y^{(l)}_t $ to forecast the next value $x_{t+1}$. Because of temporal variation, TE between $Y$ and $X$ systems is 
\begin{equation}
    TE_{Y \to X} = \sum_{x_{t+1},x^{(k)}_{t},y^{(l)}_t} p\Big( x_{t+1}, x^{(k)}_t, y^{(l)}_t\Big) \log\Bigg(\frac{p\Big(x_{t+1}\Big|x^{(k)}_t,y^{(l)}_t\Big)}{p\Big(x_{t+1}\Big|x^{(k)}_t\Big)}\Bigg).
\label{eq:2}
\end{equation}
In this way, the TE measures the deviation from the generalized Markov feature, where 
$$p\Big(x_{t+1}\Big\vert x^{(k)}_t,y^{(l)}_t\Big)$$ is the joint probability distribution of occurrence of the future $x_{t+1}$ in sync with the present states $x^{(k)}_t$ and $y^{(l)}_t$. It should be noted that, contrary to mutual information, the TE is asymmetric under the exchange of $X$ and $Y$ $:$ $T_{Y\to X}$ $\neq$ $T_{X\to Y}$ \cite{schreiber2000measuring}.

%$$
%p\Big(x_{t+1}\Big\vert x^{(k)}_t\Big):
%$$

%$$
%p \Big(x_{t+1},x^{(k)}_t, y^{(l)}_t\Big)
%$$

\subsection{Effective Transfer Entropy}

A bias may appear since finite samples are used in the TE calculation. To mitigate this bias, Sensoy \textit{et al.}  \cite{sensoy2014effective} calculated Effective Transfer Entropy (ETE), defined by,

\begin{equation}
    ETE_{Y \to X} = TE_{Y \to X}(k,l) - \frac{1}{M} \sum^{M}_{i=1}TE_{Y_{(i)}\to X^{(k,l)}} ,
    \label{eq:3}
\end{equation}
where $Y_{(i)}$ is a shuffled series of $Y$ for all $i$. Teng and Shang \cite{teng2017transfer} showed that the shuffled TE breaks the causal relationships between the variables while maintaining the probability of the distributions for each time series. ETE is calculated between TE and TE scrambled to reduce noise between TE calculations.

\section{Data}
%We analyzed the data on the daily closing value of 13  Green Bonds ETF listed in Table \ref{tab:my_label} in the period from  August 6, 2021, to October 28, 2022, collected from sites \href{https://finance.yahoo.com/}{Yahoo Finance}  and  \href{https://www.investing.com/}{Investing.com}. Was analyzed logarithmic returns: 

\begin{figure}[h]
    \centering    \includegraphics[width=14cm,height=8cm]{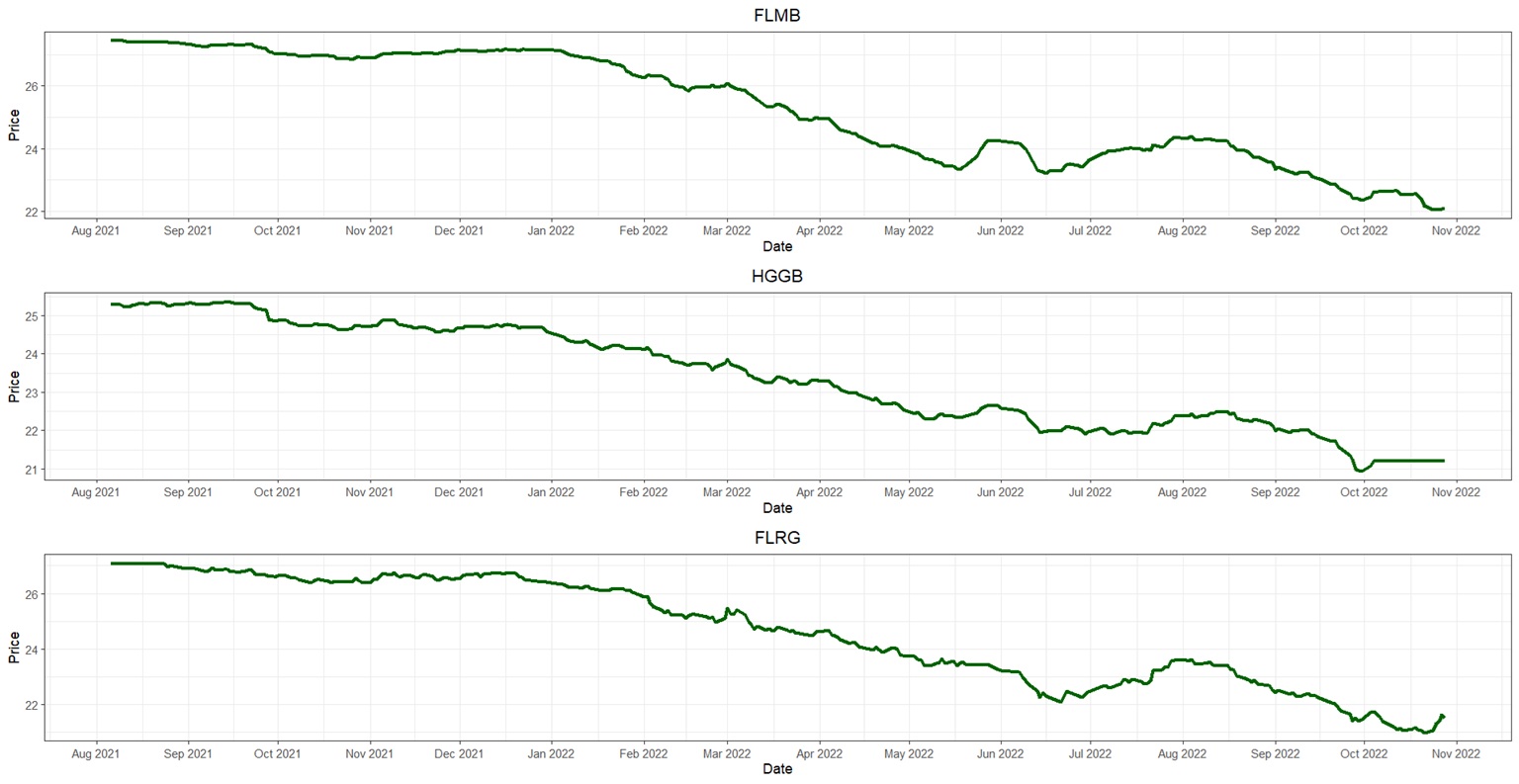}
    \caption{The graphs of the closing price series for FLMB, HGGB, and FLRG were analyzed from  August 6, 2021, to October 28, 2022.}
    \label{fig:price}
\end{figure}

%This work analyzed the daily closing price databases of 13 Green Bond ETFs listed in Table \ref{tab:my_label}times series for databases listed on \cite{harrison2023green}, from August 6, 2021, to October 28, 2022. Figure \ref{fig:price} shows the graphs of each of the 13 Green Bonds.
This work analyzed the daily closing price databases of 13 Green Bond ETFs listed in Table \ref{tab:my_label}, available in the Climate Bonds report \cite{harrison2023green}, from August 6, 2021 to October 28, 2022. %Figure \ref{fig:price} shows the charts of each of the 13 Green Bonds.
Figure \ref{fig:price} shows the time series of the Green Bonds ETF of the three markets studied in this article: FLMB (American), HGGB (Canadian), and FLRG (European).
The observation series were collected from the \href{https://finance.yahoo.com/}{Yahoo Finance} and \href{https://www.investing.com/}{Investing.com} websites. From the observation time series, the logarithmic returns for each of the ETFs were calculated using the following mathematical equation: 
\begin{equation}
    R_t = \log z_{(t)} - \log(z_{(t-1)}),\label{eq:1}
\end{equation}
where $z_{(t)}$ is the present value of the Green Bonds ETF, and $z_{(t-1)}$ is the series's one-step ago value.
\begin{table}[H]
\footnotesize
    \centering
    \caption{\textbf{List of the 13 Green Bonds ETFs analyzed in this paper.}}   
    \label{tab:my_label}
    \begin{tabular}[!]{cllc}
    \toprule
 No.& Name& Code& Currency \\
    \midrule
1 & Amundi Euro Government Green Bond UCITS ETF & EART.L & EUR\\
2& Amundi Global Aggregate Green Bond 1-10Y UCITS ETF & XCO2 & EUR\\
3 & Amundi Global Aggregate Green Bond UCITS ETF & KLMH.F & EUR \\
4 & Franklin Sustainable Euro Green Bond UCITS ETF & FLRG & EUR \\
5 & Franklin Municipal Green Bond ETF & FLMB & USD \\
6 & Horizons S\&P GreenBond Index ETF & HGGB &CAD \\
7 & L\&G ESG Green Bond UCITS ETF & GBNG.L & EUR\\
8 & iShares Global Green Bond ETF & BGRN & USA \\
9 & iShares EUR GreenBond UCITS ETF & GRON.MI & EUR\\
10 & UC MSCI European Green Bond ETF & ECBI & EUR \\
11 & Van Eck Vectors Green Bond ETF & GRNB & USA \\
12 & Xtrackers USD Corporate Green Bond UCITS ETF & XGBU.SW & USA\\
13 & Xtrackers EUR Corporate Green Bond UCITS ETF & XGBE.DE & EUR \\
\bottomrule
\end{tabular}
\end{table}

\section{RESULTS AND DISCUSSION}\label{sec:result}

\begin{figure}[H]
    \centering
\includegraphics[width=13.7cm,height=8cm]{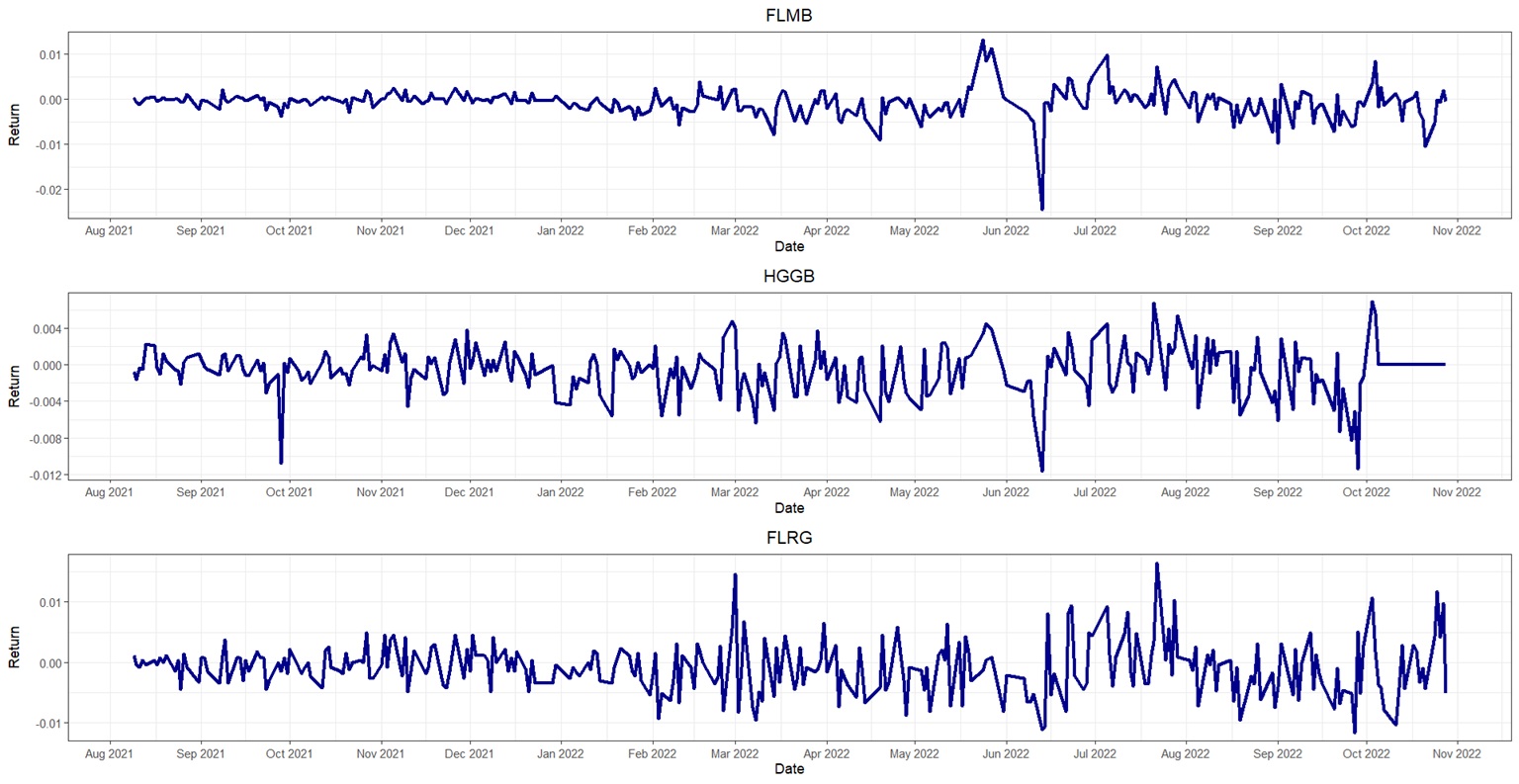}
    \caption{FLMB, HGGB and FLRG Green Bond ETF return series from  August 6, 2021, to October 28, 2022.}
    \label{fig:return}  
\end{figure}

%Table \ref{tab:table2} shows some of the descriptive statistics of the return times series for the thirteen Green Bonds ETF databases listed on \cite{harrison2023green}. As defined by Eq.(\ref{eq:1}), the period from August 6, 2021, to October 28. is shown in Figure \ref{fig:return}, where we can observe higher returns in the period.  \\
Table \ref{tab:table2} shows some descriptive statistics of the return times series for the 13 Green Bonds ETF databases listed on \cite{harrison2023green}. As defined by Eq.\ref{eq:1}, we calculate the time series of the period from August 6, 2021, to October 28, 2022. %and we can observe higher returns during that period. We also observe that in June 2022, there was a graphical change in the pattern of returns to FLMB, confirming the downward trend and change in returns.
The average variability observed in the values of these series was very similar between the series. We observed excess kurtosis in the series, especially in the FLMB series, revealing a high degree of concentration of values around its mode, in addition to heavy tails in the distribution of assets compared to a normal distribution. We also observed a higher asymmetry value in the FLMB series, which indicates a greater decline to the left of the data distribution curve. This fact may suggest that this asset is at an increased risk of large drops.
Figure \ref{fig:return} FLMB shows a smaller range of variation, showing reduced volatility and slow, HGGB shows abrupt variations and momentary suggesting greater volatility at different times, FLRG shows stable returns with few abrupt variations.
 \begin{table}[H]
\caption{\textbf{Descriptive Statistics of the Time Series of Returns of Green Bonds ETFs.}}
\centering
\begin{tabular}{lccrrr}
\toprule
 Active  & Mean & Standard Deviation & Kurtosis & Skewness \\
\midrule
BGRN & -0.00073 & 0.00442 & 3.09485 & -0.15133 \\
ECBI & -0.00090 & 0.00594 & 3.27586 & 0.18368 \\
EART.L & -0.00118 & 0.00910 & 3.27729 & 0.13516 \\
FLMB & -0.00075 & 0.00316 & 15.79349 & -1.05914 \\
FLRG & -0.00079 & 0.00427 & 4.20189 & 0.44639 \\
GBNG.L & -0.00072 & 0.00577 & 3.66154 & -0.03783 \\
GRNB & -0.00069 & 0.00379 & 4.10421 & -0.27347 \\
GRON.MI & -0.00086 & 0.00513 & 4.78791 & 0.23998 \\
HGGB & -0.00061 & 0.00268 & 5.07044 & -0.66186 \\
KLMH.F & -0.00082 & 0.00496 & 5.14898 & 0.24006 \\
XCO2 & -0.00058 & 0.00472 & 5.43667 & 0.41835 \\
XGBE.DE & -0.00071 & 0.00449 & 5.23879 & 0.36560 \\
XGBU.SW & -0.00066 & 0.00414 & 3.95108 & -0.11798 \\
\bottomrule
\end{tabular}
\label{tab:table2}
\end{table}
%Green Bonds ETF showed average values for their returns negative values in the range of  -0.00118 for minimum value and -0.00058 for maximum value, showing that the period analyzed was the moment of low for all ETFs. EART. L (European) presented the highest standard deviation. but showed values of kurtosis and asymmetry within the range of 3 to 5, which showed the highest kurtosis value was the FLMB, reaching the value of 15.79349 and a value 3 times higher than the second placed. Having no value below 3,  characterizing the data with symmetry of a normal distribution. 

All Green Bond ETFs presented negative average values for their returns, with a minimum value of $-0.00118$ and a maximum value of $-0.00058$. These values indicate that the period analyzed was a low point for all ETFs. The Green Bound EART.L (European) presented the highest standard deviation but presented kurtosis and asymmetry values close to those of a normal distribution. The highest kurtosis value was $15.79349$, reached by Green Bound FLMB. This value is almost 3 times higher than the second-highest (XCO2 with a kurtosis of 5.43667). Among the observed databases, no kurtosis lower than three was found,  characterizing the distribution curve behavior as very close to that of a Gaussian distribution.
%Figure \ref{fig:return} shows that in June 2022, there was a graphical change in the pattern of returns to FLMB, confirming the downward trend and change in returns.  

%We applied Shannon's Transfer Entropy and Effective Transfer Entropy to quantify the flow directional information among Green Bond ETFS. Table \ref{tab:cae} exhibit the values of transfer entropy from Canadian and European ETFs, Table \ref{tab:caa} exhibit the values of transfer entropy from Canadian and American ETFs, Table \ref{tab:american} exhibit the values of transfer entropy from  American EFT,  Table \ref{tab:European} exhibit the values of transfer entropy from European ETFs, Table \ref{tab:EuropeanandAmerican} exhibit the values of transfer entropy from European and American ETFs in alphabetically.
We applied Transfer Entropy and Effective Transfer Entropy to quantify the flow direction information among Green Bond ETFS. Tables \ref{tab:cae} to \ref{tab:EuropeanandAmerican} show the results obtained for the transfer entropy values:
\begin{itemize}
    \item between Canadian and European ETFs (Table \ref{tab:cae});
    \item between Canadian and American ETFs (Table \ref{tab:caa});
    \item only American ETFs (Table \ref{tab:american});
    \item only European ETFs (Table \ref{tab:European});
    \item between European and American ETFs (Table \ref{tab:EuropeanandAmerican}).
\end{itemize}

%In which the tables (*) represent 
For all results of $p-$values, the notation with marks was used: (.) to represent a significance level of $0.1\%$, (*) to represent a significance level of $0.05\%$, (**) to represent a significance level of $0.01\%$, and (***) to represent a significance level of $0.001\%$. The absence of marks (asterisks or dots) means a lack of statistical significance within the established level minimum of $0.001$.

Table \ref{tab:cae} shows that the information transfer between ETFs is not symmetric between the Canadian and European markets. %In many cases, There is no information interchange, being often unidimensional. 
In many cases, there is no information exchange as the transfer is often one-directional.
The Canadian ETF HGGB exerts a significant statistical influence on the FLRG: TE = 0.0401 and ETE = 0.0207, KLMH.F ETFs: TE = 0.0562 and ETE = 0.0380, EART.L: TE = 0.0472 and ETE = 0.0306, XCO2: TE = 0.0428 and ETE = 0.0253, GRON.MI: TE =  0.0341 and ETE = 0.0186 , GBNG.L: TE = 0.0329 and ETE = 0.0177 and XGBE.DE: TE = 0.0341 and ETE = 0.0145. However, the inverse influence is not statistically significant for all cases.
The HGGB ETF does not receive statistically significant information from the FLRG and EART.L ETFs, ECBI, and XCO2. The HGGB ETF receives information from GRON.MI: TE = 0.0355 and ETE = 0.0186 ($0.0067$***). This fact suggests a strong interactive relationship. Highlighting that the KLMH.F ETF: TE = 0.0562 and ETE = 0.0380 receive the largest transfer of information from HGGB with a $p-$value of $0.0000$***. The ETF XGBE.DE: TE = 0.0491 and ETE = 0.0329 strongly influences the HGGB ETF with a $p-$value of $0.0000$***, indicating that the Canadian market may also be relevant to the European market. %$and the performance of kurtosis as a high transfer index. 
\begin{table}[H]
\centering
\caption{\textbf{TE and ETE results among Canadian and European ETFs.}}
\label{tab:cae}
\begin{tabular}{lcccl}
\toprule
Direction & TE & ETE & Std. Err. & \textit{p}-value \\
\midrule
EART.L→HGGB & 0.0222 & 0.0069 & 0.0062 & 0.1333 \\
HGGB→EART.L & 0.0472 & 0.0306 & 0.0080 & 0.0067** \\
ECBI→HGGB & 0.0086 & 0.0000 & 0.0067 & 0.8900 \\
HGGB→ECBI & 0.0349 & 0.0169 & 0.0087 & 0.0500 \\
FLRG→HGGB & 0.0177 & 0.0016 & 0.0075 & 0.2867 \\
HGGB→FLRG & 0.0401 & 0.0207 & 0.0090 & 0.0233* \\
GBNG.L→HGGB & 0.0307 & 0.0142 & 0.0067 & 0.0267* \\
HGGB→GBNG.L & 0.0329 & 0.0177 & 0.0066 & 0.0100* \\
GRON.MI→HGGB & 0.0355 & 0.0186 & 0.0065 & 0.0067** \\
HGGB→GRON.MI & 0.0341 & 0.0189 & 0.0075 & 0.0300* \\
KLMH.F→HGGB & 0.0245 & 0.0097 & 0.0066 & 0.1100 \\
HGGB→KLMH.F & 0.0562 & 0.0380 & 0.0085 & 0.0000*** \\
XCO2→HGGB & 0.0174 & 0.0021 & 0.0072 & 0.2933 \\
HGGB→XCO2 & 0.0428 & 0.0253 & 0.0081 & 0.0167* \\
XGBE.DE→HGGB & 0.0491 & 0.0329 & 0.0072 & 0.0000*** \\
HGGB→XGBE.DE & 0.0341 & 0.0145 & 0.0069 & 0.0267* \\
\bottomrule
\end{tabular}
\end{table}

\begin{figure}[H]
    \centering
    \begin{subfigure}{0.45\textwidth}
        \centering
        \includegraphics[width=\linewidth]{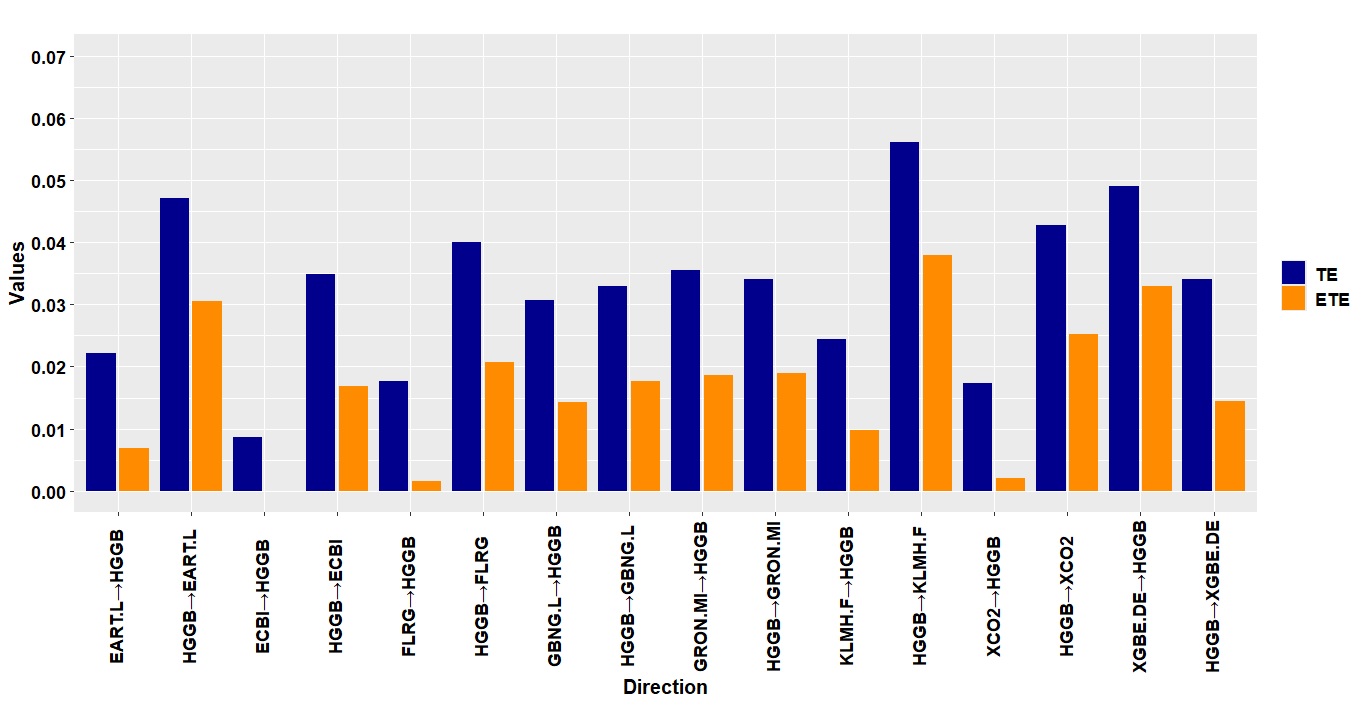}
        \caption{TE and ETE results among Canadian and European ETFs.}
        \label{fig:Rplot01}
    \end{subfigure}
    \hspace{0.03\textwidth} 
    \begin{subfigure}{0.45\textwidth}
        \centering
        \includegraphics[width=\linewidth]{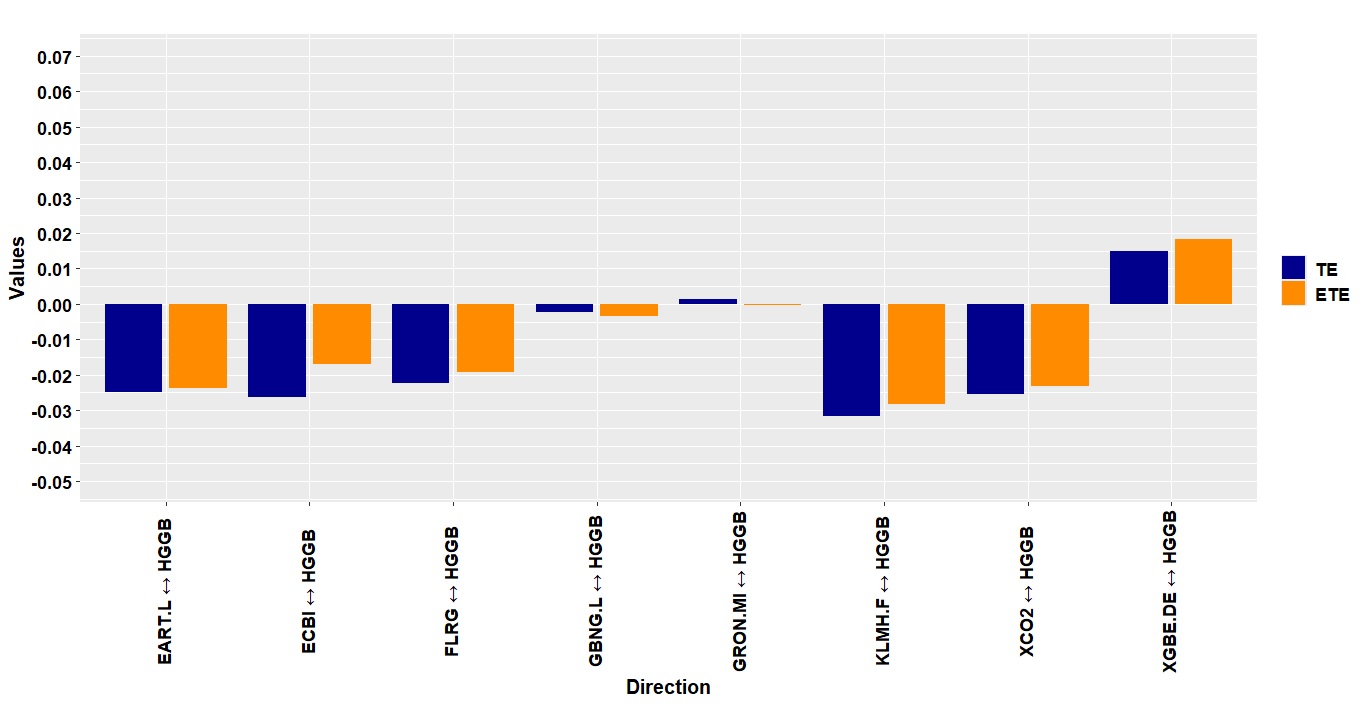}
        \caption{TE and ETE difference among Canadian and European ETFs.}
        \label{fig:Rplot011}
    \end{subfigure}
    \hspace{0.03\textwidth}
     \caption{TE and ETE results and difference among Canadian and European ETF.}
    \end{figure}

 The results of Table \ref{tab:cae} are presented in Figure \ref{fig:Rplot01}, where one can observe the asymmetry in the direction of the information flow. The difference between the Canadian and European ETFs shown in Figure \ref{fig:Rplot011} with positive values indicates that the HGGB ETF receives more information. However, negative values appear when it receives more information than it sends. %The difference between the Canadian and European ETFs shown in Figure \ref{fig:Rplot011} positive values indicate that the HGGB ETF sends more information and negative values when it receives more information than it sends. 
 
\begin{figure}[H]
    \centering
    \begin{subfigure}{0.45\textwidth}
        \centering
        \includegraphics[width=\linewidth]{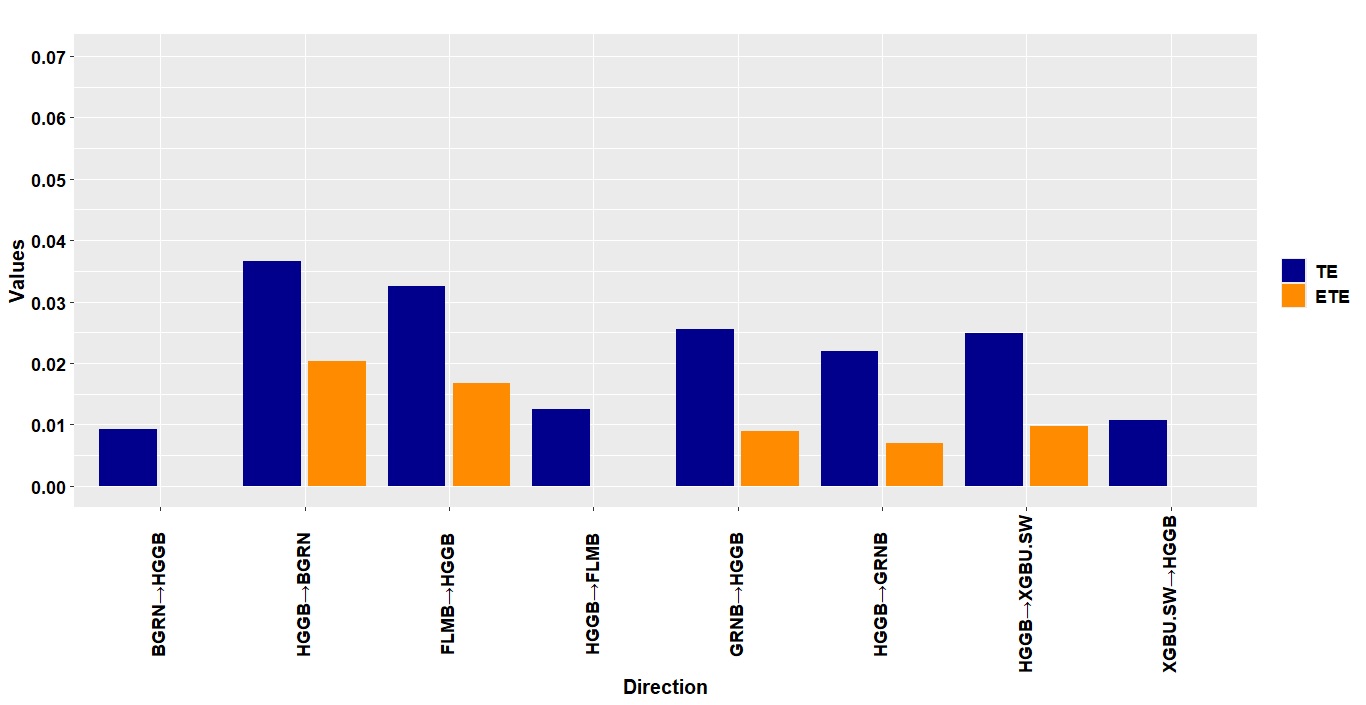}
        \caption{TE and ETE results among Canadian and American ETFs.}
        \label{fig:Rplot00}
    \end{subfigure}
    \hspace{0.03\textwidth} 
    \begin{subfigure}{0.45\textwidth}
      \centering
        \includegraphics[width=\linewidth]{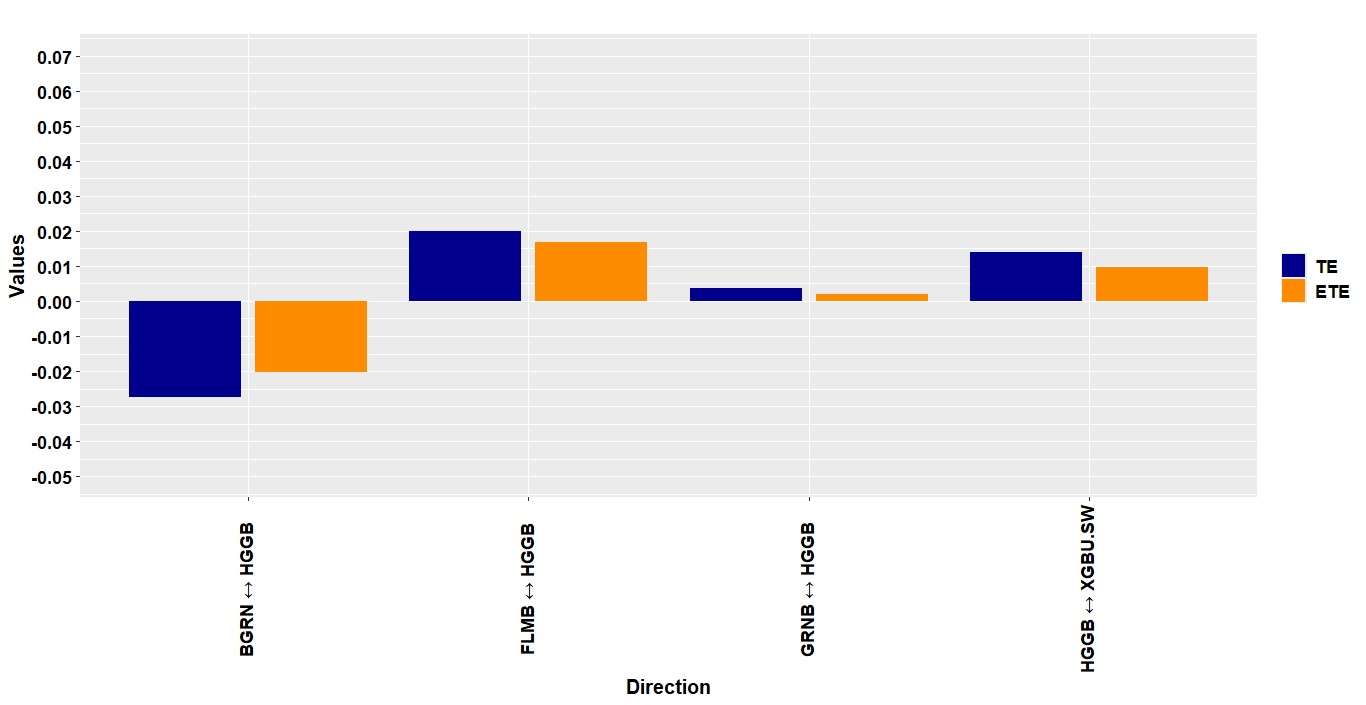}
        \caption{TE and ETE  difference among Canadian and American ETFs.}
        \label{fig:Rplot000}   
    \end{subfigure}
    \hspace{0.03\textwidth}
    \caption{TE and ETE results and difference among Canadian and American ETF.}
    \end{figure}

%\begin{table}[H]
   % \centering
    %\caption{\textbf{Results among Canadian and American ETFs.}}
   % \label{tab:caa}
    %\begin{tabular}{lcccl}
    %\toprule
    % Direction & TE & ETE & Std. Err. & \textit{p}Value  \\
   % \midrule
      %  FLMB→HGGB & 0.0325 & 0.0167 & 0.0063 & 0.0167* \\
       % HGGB→FLMB & 0.0125 & 0.0000 & 0.0074 & 0.5300 \\
       % HGGB→BGRN & 0.0366 & 0.0204 & 0.0085 & 0.0233* \\
       % BGRN→HGGB & 0.0093 & 0.0000 & 0.0078 & 0.8467 \\
       % HGGB→GRNB & 0.0219 & 0.0069 & 0.0065 & 0.1167 \\
       % GRNB→HGGB & 0.0255 & 0.0089 & 0.0069 & 0.0933 \\
       % HGGB→XGBU.SW & 0.0249 & 0.0097 & 0.0072 & 0.1233 \\
       % XGBU.SW→HGGB & 0.0108 & 0.0000 & 0.0075 & 0.7767 \\
   % \bottomrule
   % \end{tabular}
%\end{table}
Table \ref{tab:caa} shows that the transfer of information between ETFs in the Canadian and USA markets is also not symmetric. HGGB: TE = 0.0366, ETE = 0.0204 influences the BGRN ETF statistically significantly ($p-$value of $0.0233$* ). In contrast, the ETF FLMB: TE = 0.0325 and  ETE = 0.0167 considerably influence HGGB, with a p-value of 0.0167*. Other ETFs do not influence or exert a statistically significant influence. The HGGB has greater influence on American ETFs, so we can see that there is evidence of a limited influence of the American market in the Canadian market because the only American ETF that could send information to the Canadian HGGB  ETF was the FMLB.
\begin{table}[H]
    \centering
    \caption{\textbf{TE and ETE results among Canadian and American ETFs.}}
    \label{tab:caa}
    \begin{tabular}{lcccl}
    \toprule
     Direction & TE & ETE & Std. Err. & \textit{p}-value  \\
    \midrule
        BGRN→HGGB & 0.0093 & 0.0000 & 0.0078 & 0.8467 \\
        HGGB→BGRN & 0.0366 & 0.0204 & 0.0085 & 0.0233* \\
        FLMB→HGGB & 0.0325 & 0.0167 & 0.0063 & 0.0167* \\
        HGGB→FLMB & 0.0125 & 0.0000 & 0.0074 & 0.5300 \\
        GRNB→HGGB & 0.0255 & 0.0089 & 0.0069 & 0.0933 \\
        HGGB→GRNB & 0.0219 & 0.0069 & 0.0065 & 0.1167 \\
        HGGB→XGBU.SW & 0.0249 & 0.0097 & 0.0072 & 0.1233 \\
        XGBU.SW→HGGB & 0.0108 & 0.0000 & 0.0075 & 0.7767 \\
    \bottomrule
    \end{tabular}
\end{table}
The results of Table \ref{tab:caa} shown in Figure \ref{fig:Rplot00} show asymmetry in the information flow. In Figure \ref{fig:Rplot000}, a balance between the sending of information between the Canadian and American ETFs is seen in that of the four pairs of TE and ETE,  two the Canadian ETF sends information, and two receives differently than in Figure \ref{fig:Rplot011},  where only one European EFT sent more information than it received from Canada. 
    
%The results of Table \ref{tab:caa} presented in Figure \ref{fig:Rplot00} show asymmetry in the flow of information. In Figure \ref{fig:Rplot000} it is seen that the positive values are mixed between a greater sending of information from ETF HGGB and receiving information differently than in Figure \ref{fig:Rplot011} where there were only positive values when sending more information than received.%, in this figure showing negative value only when the HGGB received more information from GRMB.

 %\begin{table}[H]
%\centering
%\caption{\textbf{Results among American EFTs.}}
%\label{tab:american}
%\begin{tabular}{lcccl}
%\toprule
%Direction & TE & ETE & Std. Err. & \textit{p}-Value \\
%\midrule
%BGRN→FLMB & 0.0144 & 0.0000 & 0.0064 & 0.4267 \\
%FLMB→BGRN & 0.0452 & 0.0275 & 0.0080 & 0.0033** \\
%FLMB→GRNB & 0.0351 & 0.0200 & 0.0082 & 0.0267* \\
%GRNB→FLMB & 0.0165 & 0.0006 & 0.0072 & 0.3767 \\
%FLMB→XGBU.SW & 0.0221 & 0.0068 & 0.0070 & 0.1500 \\
%XGBU.SW→FLMB & 0.0091 & 0.0000 & 0.0062 & 0.8300 \\
%GRNB→BGRN & 0.0137 & 0.0000 & 0.0082 & 0.5633 \\
%BGRN→GRNB & 0.0129 & 0.0000 & 0.0074 & 0.5133 \\
%XGBU.SW→BGRN & 0.0226 & 0.0065 & 0.0079 & 0.2100 \\
%BGRN→XGBU.SW & 0.0196 & 0.0042 & 0.0075 & 0.2333 \\
%XGBU.SW→GRNB & 0.0133 & 0.0000 & 0.0067 & 0.4533 \\
%GRNB→XGBU.SW & 0.0273 & 0.0104 & 0.0067 & 0.0733 \\
%\bottomrule
%\end{tabular}
%\end{table}   

 Table \ref{tab:american} shows that the American market does not have a symmetric information transfer, with the FLMB ETF as the main transmitter of information for the ETFs: FLMB for BGRN has TE = 0.0452 and  ETE = 0.0275 ($p-$value of $0.0033$**) and FLMB has TE = 0.0351 and  ETE = 0.0200 ($p-$value of $0.0267$*) to GRNB; indicating that it has a strong influence on these ETFs, being a one-dimensional transfer because they do not send statistically significant information. Therefore, the FLMB ETF is considered dominant over other ETFs. The other ETFs did not show statistically significant transfers. 

\begin{figure}[H]
    \centering
    \begin{subfigure}[t]{0.45\textwidth} 
        \centering
        \includegraphics[width=\linewidth, height=0.25\textheight]{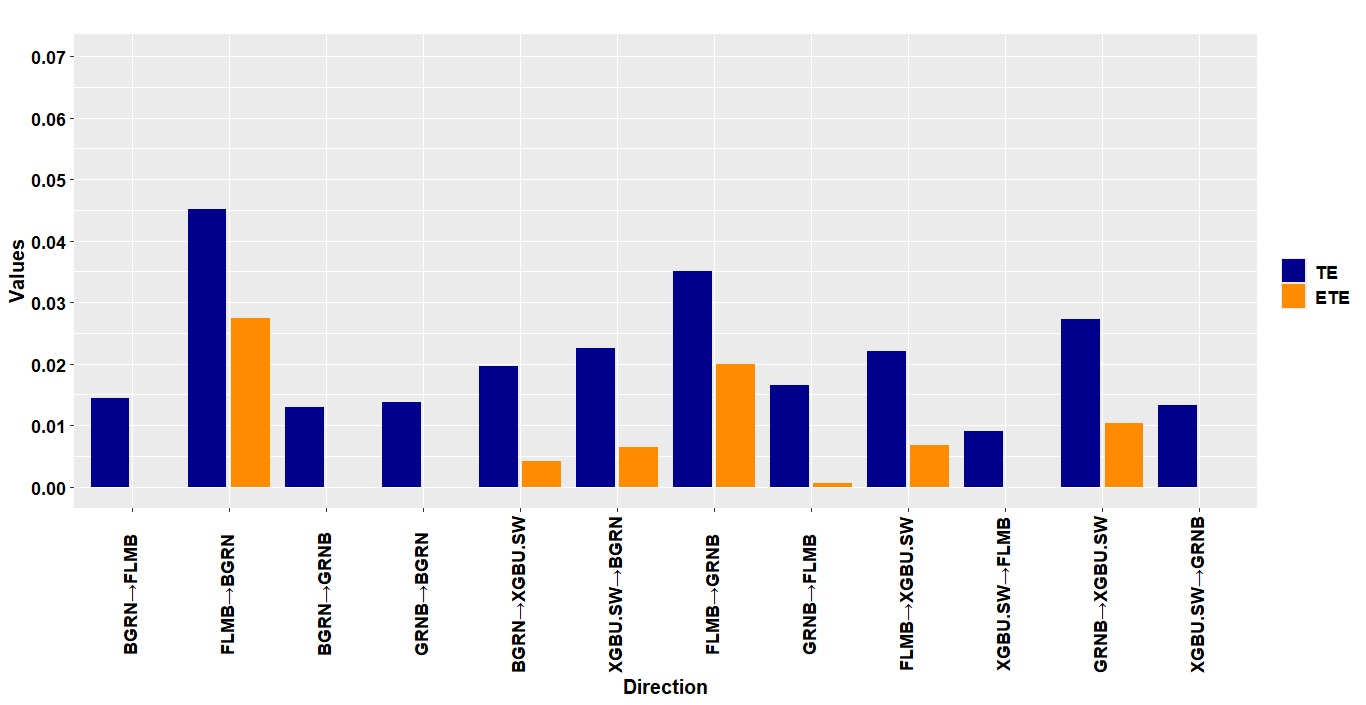}
        \caption{TE and ETE results among American ETFs.}
        \label{fig:Rplot00american}
    \end{subfigure}
    \hspace{0.03\textwidth} 
    \begin{subfigure}[t]{0.45\textwidth} 
        \centering
        \includegraphics[width=\linewidth, height=0.25\textheight]{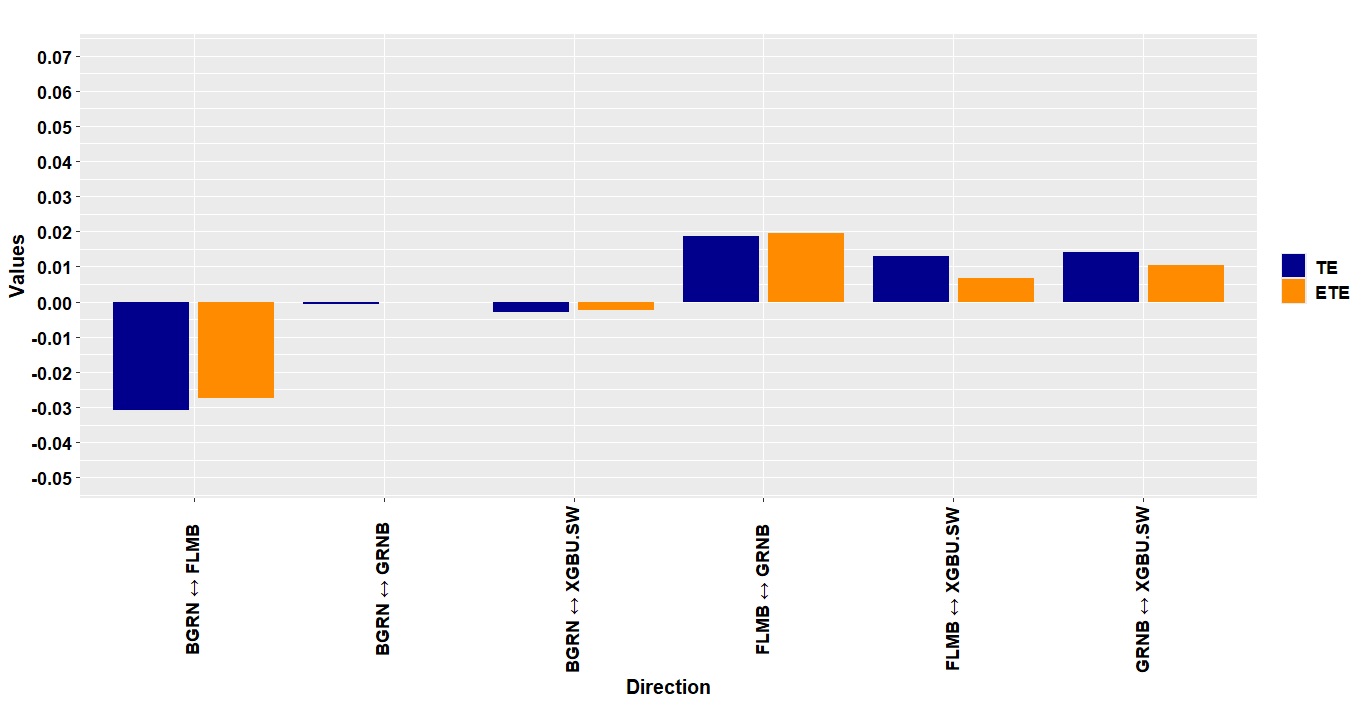} % Ajusta largura e altura da imagem
        \caption{TE and ETE difference among American ETFs.}
        \label{fig:Rplot01diference}
    \end{subfigure}
    \caption{TE and ETE results and difference among American ETFs.}
    \label{fig:shannon_te}
\end{figure}

\begin{table}[H]
\centering
\caption{\textbf{TE and ETE results among American ETFs.}}
\label{tab:american}
\begin{tabular}{lcccl}
\toprule
Direction & TE & ETE & Std. Err. & \textit{p}-value \\
\midrule
BGRN→FLMB & 0.0144 & 0.0000 & 0.0064 & 0.4267 \\
FLMB→BGRN & 0.0452 & 0.0275 & 0.0080 & 0.0033** \\
BGRN→GRNB & 0.0129 & 0.0000 & 0.0074 & 0.5133 \\
GRNB→BGRN & 0.0137 & 0.0000 & 0.0082 & 0.5633 \\
BGRN→XGBU.SW & 0.0196 & 0.0042 & 0.0075 & 0.2333 \\
XGBU.SW→BGRN & 0.0226 & 0.0065 & 0.0079 & 0.2100 \\
FLMB→GRNB & 0.0351 & 0.0200 & 0.0082 & 0.0267* \\
GRNB→FLMB & 0.0165 & 0.0006 & 0.0072 & 0.3767 \\
FLMB→XGBU.SW & 0.0221 & 0.0068 & 0.0070 & 0.1500 \\
XGBU.SW→FLMB & 0.0091 & 0.0000 & 0.0062 & 0.8300 \\
GRNB→XGBU.SW & 0.0273 & 0.0104 & 0.0067 & 0.0733 \\
XGBU.SW→GRNB & 0.0133 & 0.0000 & 0.0067 & 0.4533 \\
\bottomrule
\end{tabular}
\end{table}
The results of Table \ref{tab:american} presented in Figure \ref{fig:Rplot00american} reveal the asymmetry in the flow of information even if the ETFs are from the same country or continent. The difference shown in Figure \ref{fig:Rplot01diference} shows the FLMB ETF as a major information transmitter among ETFs. The negative values were the responsibility of ETF BGRN, which received more information than it sent.

\begin{figure}[H]
  \centering
    \begin{subfigure}{0.80\textwidth}
        \centering
        \includegraphics[width=\linewidth]{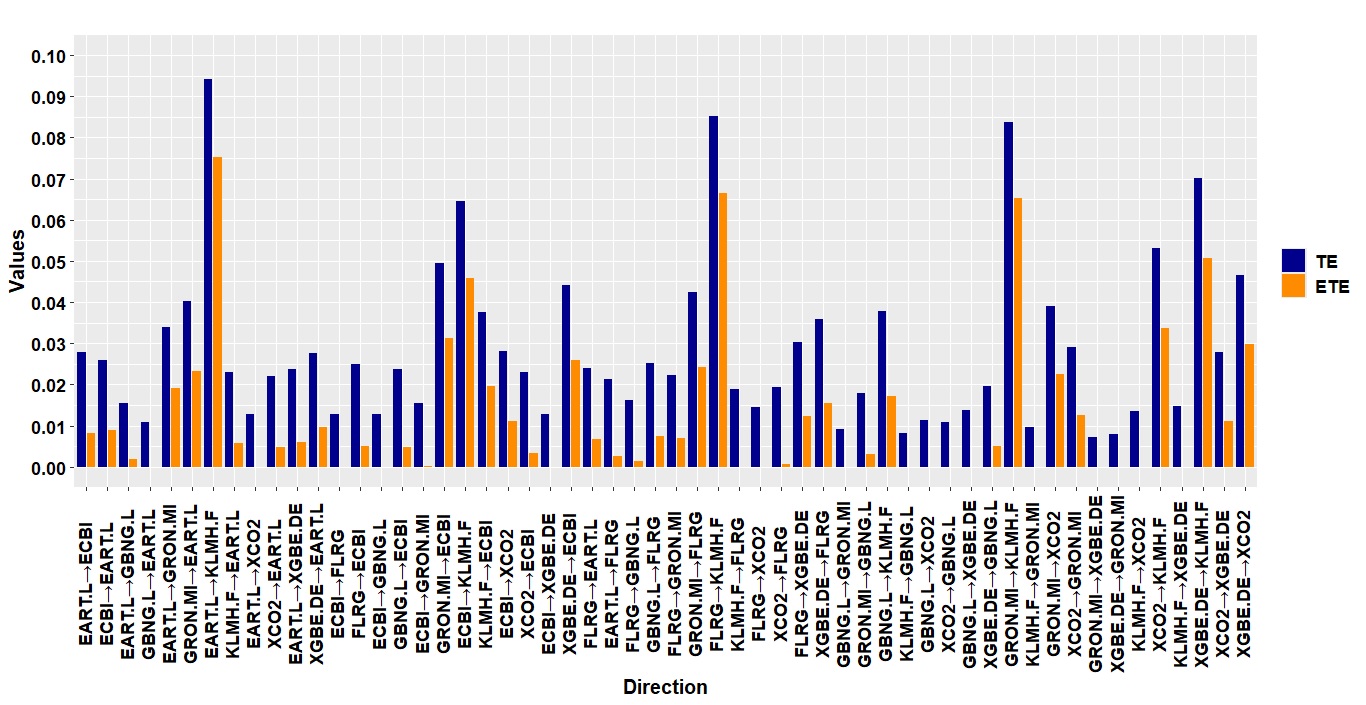} 
     \caption{TE and ETE results among European ETFs.}
     \label{fig:Rplot02}
    \end{subfigure}
        \label{fig:european}
    \begin{subfigure}{0.80\textwidth}
 \centering
        \includegraphics[width=\linewidth]{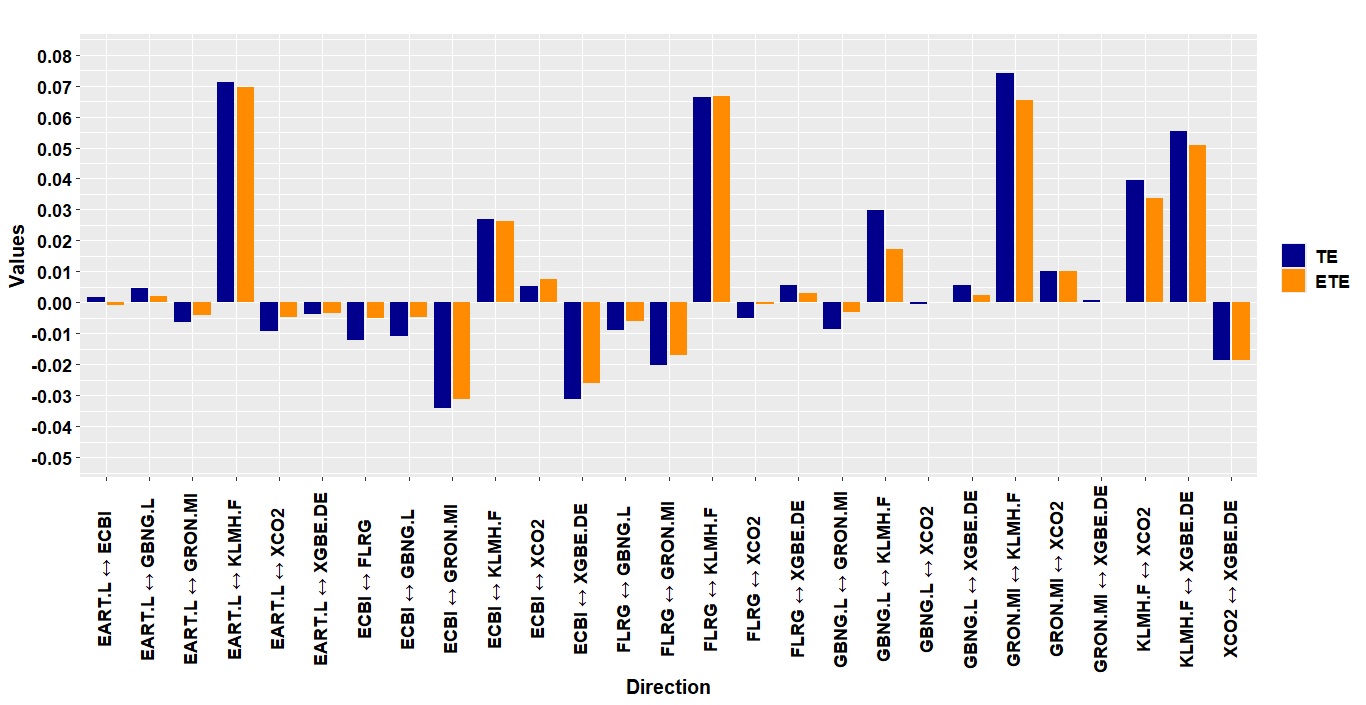} 
     \caption{TE and ETE difference among European ETFs.}
     \label{fig:european10}
    \end{subfigure}
      \caption{TE and ETE results and difference among  European ETFs.}
\end{figure}

 Table \ref{tab:European} exhibits that the transfer of information in the European market is not symmetrical, so we do not have a two-way transfer between ETFs, showing that some funds are more influential than others. Among the European funds, the importance of ETFs: FLRG and GRON.MI. In particular, GRON.MI is observed as a great source of information in the market, sending a lot of information, especially to the ETFs KLMH.F and EART.L. The ETF GRON.MI: TE = 0.0425 and  ETE = 0.0242 sends the largest amount of information in table to FLRG ($p-$value of 0.0133*); the ETF KLMH.F receives information from almost all other ETFs, particularly from EART.L receives TE = 0.0942 and  ETE =  0.0753 ($p-$value of $0.0000$***) of information and second receiving from FLRG TE = 0.0852 and  ETE = 0.0666 ($p-$value of $0.0000$***), and not only does it receive, but also sends information with statistical significance to the KLMH.F for ETF ECBI  TE = 0.0377 and  ETE = 0.0195 ($p-$value of $0.0467$*). Within the FLRG ETFs, it stands out as a major information issuer along with GRON.MI and EART. L, the GRON.MI sends a lot of information; however does not receive as much information, being influenced in a specific way within the European market. KLMH.F shows a point of centralization of information in the European market.
\begin{longtable}{lcccl}
\caption{\textbf{TE and ETE results among European ETFs}}
\label{tab:European}\\
\hline
Direction & TE & ETE & Std. Err. & \textit{p}-value \\
\hline
\endfirsthead
\hline
Direction & TE & ETE & Std. Err. & \textit{p}-value \\
\hline
\endhead
\hline
\multicolumn{5}{r}{Continued on next page} \\
\endfoot
\hline
\endlastfoot
EART.L→ECBI & 0.0278 & 0.0081 & 0.0082 & 0.1033 \\
ECBI→EART.L & 0.0260 & 0.0089 & 0.0085 & 0.1233 \\
EART.L→GBNG.L & 0.0155 & 0.0020 & 0.0071 & 0.2800 \\
GBNG.L→EART.L & 0.0109 & 0.0000 & 0.0082 & 0.7067 \\
EART.L→GRON.MI & 0.0340 & 0.0192 & 0.0075 & 0.0233* \\
GRON.MI→EART.L & 0.0403 & 0.0232 & 0.0087 & 0.0233* \\
EART.L→KLMH.F & 0.0942 & 0.0753 & 0.0083 & 0.0000*** \\
KLMH.F→EART.L & 0.0230 & 0.0058 & 0.0074 & 0.1433 \\
EART.L→XCO2 & 0.0128 & 0.0000 & 0.0078 & 0.6067 \\
XCO2→EART.L & 0.0220 & 0.0048 & 0.0081 & 0.1733 \\
EART.L→XGBE.DE & 0.0238 & 0.0061 & 0.0083 & 0.1900 \\
XGBE.DE→EART.L & 0.0277 & 0.0096 & 0.0074 & 0.0867 \\
ECBI→FLRG & 0.0129 & 0.0000 & 0.0087 & 0.6133 \\
FLRG→ECBI & 0.0249 & 0.0050 & 0.0088 & 0.1800 \\
ECBI→GBNG.L & 0.0129 & 0.0000 & 0.0063 & 0.4067 \\
GBNG.L→ECBI & 0.0238 & 0.0048 & 0.0091 & 0.2400 \\
ECBI→GRON.MI & 0.0154 & 0.0001 & 0.0071 & 0.3800 \\
GRON.MI→ECBI & 0.0494 & 0.0313 & 0.0094 & 0.0100* \\
ECBI→KLMH.F & 0.0646 & 0.0458 & 0.0085 & 0.0000*** \\
KLMH.F→ECBI & 0.0377 & 0.0195 & 0.0096 & 0.0467* \\
ECBI→XCO2 & 0.0282 & 0.0110 & 0.0078 & 0.0733 \\
XCO2→ECBI & 0.0229 & 0.0034 & 0.0088 & 0.2400 \\
ECBI→XGBE.DE & 0.0129 & 0.0000 & 0.0082 & 0.6300 \\
XGBE.DE→ECBI & 0.0441 & 0.0260 & 0.0077 & 0.0067** \\
FLRG→EART.L & 0.0239 & 0.0068 & 0.0078 & 0.1700 \\
EART.L→FLRG & 0.0214 & 0.0026 & 0.0081 & 0.2567 \\
FLRG→GBNG.L & 0.0162 & 0.0015 & 0.0069 & 0.2633 \\
GBNG.L→FLRG & 0.0251 & 0.0075 & 0.0080 & 0.1533 \\
FLRG→GRON.MI & 0.0223 & 0.0071 & 0.0076 & 0.1767 \\
GRON.MI→FLRG & 0.0425 & 0.0242 & 0.0079 & 0.0133* \\
FLRG→KLMH.F & 0.0852 & 0.0666 & 0.0079 & 0.0000*** \\
KLMH.F→FLRG & 0.0190 & 0.0000 & 0.0102 & 0.4133 \\
FLRG→XCO2 & 0.0145 & 0.0000 & 0.0081 & 0.4500 \\
XCO2→FLRG & 0.0194 & 0.0006 & 0.0086 & 0.3700 \\
FLRG→XGBE.DE & 0.0303 & 0.0123 & 0.0081 & 0.0567. \\
XGBE.DE→FLRG & 0.0358 & 0.0154 & 0.0086 & 0.0333* \\
GBNG.L→GRON.MI & 0.0092 & 0.0000 & 0.0072 & 0.8400 \\
GRON.MI→GBNG.L & 0.0178 & 0.0031 & 0.0073 & 0.2300 \\
GBNG.L→KLMH.F & 0.0378 & 0.0173 & 0.0099 & 0.0433* \\
KLMH.F→GBNG.L & 0.0081 & 0.0000 & 0.0063 & 0.7533 \\
GBNG.L→XCO2 & 0.0114 & 0.0000 & 0.0085 & 0.6967 \\
XCO2→GBNG.L & 0.0109 & 0.0000 & 0.0065 & 0.5600 \\
GBNG.L→XGBE.DE & 0.0139 & 0.0000 & 0.0073 & 0.5333 \\
XGBE.DE→GBNG.L & 0.0195 & 0.0051 & 0.0067 & 0.1700 \\
GRON.MI→KLMH.F & 0.0837 & 0.0654 & 0.0084 & 0.0000*** \\
KLMH.F→GRON.MI & 0.0096 & 0.0000 & 0.0069 & 0.7800 \\
GRON.MI→XCO2 & 0.0391 & 0.0226 & 0.0078 & 0.0167* \\
XCO2→GRON.MI & 0.0290 & 0.0126 & 0.0072 & 0.0533. \\
GRON.MI→XGBE.DE & 0.0073 & 0.0000 & 0.0073 & 0.9133 \\
XGBE.DE→GRON.MI & 0.0079 & 0.0000 & 0.0073 & 0.9133 \\
KLMH.F→XCO2 & 0.0136 & 0.0000 & 0.0092 & 0.5500 \\
XCO2→KLMH.F & 0.0532 & 0.0338 & 0.0093 & 0.0033** \\
KLMH.F→XGBE.DE & 0.0147 & 0.0000 & 0.0073 & 0.4900 \\
XGBE.DE→KLMH.F & 0.0701 & 0.0508 & 0.0093 & 0.0000*** \\
XCO2→XGBE.DE & 0.0278 & 0.0110 & 0.0082 & 0.1167 \\
XGBE.DE→XCO2 & 0.0465 & 0.0297 & 0.0083 & 0.0100* \\
\end{longtable}
Figure \ref{fig:Rplot02} showing the results of Table \ref{tab:European} reveals the asymmetry in the information flow among the European ETFs, corroborating with what was observed in Figure \ref{fig:Rplot00american} that ETFs from the same locality did not show symmetry in the information flow. Figure \ref{fig:european10} shows the difference in the flow of information from European ETFs, the positive values were mostly ETF KLMH.F receiving information from other ETFs, if showing an ETFs with great ability to receive information, already the negative values were mostly the ETF KLMH.F sending information, showing its low capacity to transmit information.

Table \ref{tab:EuropeanandAmerican} exhibits that the information transfers between the European and American markets also show an asymmetric market in which the FLRG and FLMB ETFs have no correlated values. Still, FLMB TE = 0.0411 and  ETE = 0.0231 ($p-$value of $0.0200$*) sends more information to FLRG than the opposite. FLMB is the ETF that transfers the most information between funds, but FLRG and BGRN are also great for forecasting the movement of other ETFs. FLMB TE = 0.0598, ETE = 0.0407 ($p-$value of $0.0000$***) sends a strong information transfer to KLMH.F, FLMB:  TE = 0.0319 and  ETE = 0.0154 sends more information to GRON.MI ($p-$value of $0.0267$*), demonstrating that FLMB ETF plays a central role in the transmission of information both within the American market and through European ETFs
FLRG transfers large amounts of information to BGRN with TE = 0.0472 and ETE = 0.0298 ($p-$value of $0.0067$**) and GRNB with TE = 0.0413 and ETE = 0.0275 ($p-$value of $0.0033$**). BGRN TE = 0.0452 and ETE = 0.0287 transfer a lot of information to XCO2 ($p-$value of $0.0000$***). With the prevalence of BGRN and FLRG ETFs, it is perceived that the American market predominates in transferring information with greater intensity to the European market.
\begin{longtable}{lcccl}

\caption{\textbf{Shannon TE results among European and American ETFs}}
\label{tab:EuropeanandAmerican}\\
\hline
Direction & TE & ETE & Std. Err. & \textit{p}-Value \\
\hline
\endfirsthead
\hline
Direction & TE & ETE & Std. Err. & \textit{p}-Value \\
\hline
\endhead
\hline
\multicolumn{5}{r}{Continued on next page} \\
\endfoot
\hline
\endlastfoot
BGRN→EART.L & 0.0174 & 0.0015 & 0.0083 & 0.3867 \\
EART.L→BGRN & 0.0186 & 0.0000 & 0.0088 & 0.4133 \\
BGRN→ECBI & 0.0303 & 0.0116 & 0.0084 & 0.0767. \\
ECBI→BGRN & 0.0232 & 0.0058 & 0.0076 & 0.1933 \\
BGRN→GBNG.L & 0.0294 & 0.0144 & 0.0070 & 0.0500. \\
GBNG.L→BGRN & 0.0378 & 0.0200 & 0.0081 & 0.0300* \\
BGRN→GRON.MI & 0.0232 & 0.0067 & 0.0071 & 0.1367 \\
GRON.MI→BGRN & 0.0157 & 0.0000 & 0.0082 & 0.4833 \\
BGRN→KLMH.F & 0.0564 & 0.0375 & 0.0085 & 0.0000*** \\
KLMH.F→BGRN & 0.0213 & 0.0029 & 0.0084 & 0.2533 \\
BGRN→XCO2 & 0.0452 & 0.0287 & 0.0078 & 0.0000*** \\
XCO2→BGRN & 0.0299 & 0.0123 & 0.0092 & 0.0900. \\
BGRN→XGBE.DE & 0.0226 & 0.0056 & 0.0082 & 0.2167 \\
XGBE.DE→BGRN & 0.0167 & 0.0003 & 0.0090 & 0.4200 \\
BGRN→FLRG & 0.0374 & 0.0192 & 0.0082 & 0.0233* \\
FLRG→BGRN & 0.0472 & 0.0298 & 0.0082 & 0.0067** \\
EART.L→FLMB & 0.0084 & 0.0000 & 0.0066 & 0.8767 \\
FLMB→EART.L & 0.0292 & 0.0119 & 0.0076 & 0.0567 \\
EART.L→GRNB & 0.0124 & 0.0000 & 0.0069 & 0.4900 \\
GRNB→EART.L & 0.0162 & 0.0000 & 0.0077 & 0.3533 \\
EART.L→XGBU.SW & 0.0108 & 0.0000 & 0.0075 & 0.7100 \\
XGBU.SW→EART.L & 0.0372 & 0.0215 & 0.0077 & 0.0200 * \\
ECBI→FLMB & 0.0112 & 0.0000 & 0.0064 & 0.6800 \\
FLMB→ECBI & 0.0279 & 0.0090 & 0.0084 & 0.1267 \\
ECBI→GRNB & 0.0219 & 0.0068 & 0.0064 & 0.1100 \\
GRNB→ECBI & 0.0240 & 0.0031 & 0.0084 & 0.2500 \\
ECBI→XGBU.SW & 0.0142 & 0.0000 & 0.0071 & 0.3967 \\
XGBU.SW→ECBI & 0.0161 & 0.0000 & 0.0088 & 0.4633 \\
FLMB→GBNG.L & 0.0210 & 0.0060 & 0.0067 & 0.1000 \\
GBNG.L→FLMB & 0.0091 & 0.0000 & 0.0063 & 0.7733 \\
FLMB→GRON.MI & 0.0319 & 0.0154 & 0.0070 & 0.0267* \\
GRON.MI→FLMB & 0.0099 & 0.0000 & 0.0068 & 0.7933 \\
FLMB→KLMH.F & 0.0598 & 0.0407 & 0.0076 & 0.0000*** \\
KLMH.F→FLMB & 0.0231 & 0.0082 & 0.0078 & 0.0900. \\
FLMB→XCO2 & 0.0340 & 0.0166 & 0.0093 & 0.0500 \\
XCO2→FLMB & 0.0155 & 0.0007 & 0.0083 & 0.4067 \\
FLMB→XGBE.DE & 0.0270 & 0.0091 & 0.0079 & 0.0967 \\
XGBE.DE→FLMB & 0.0107 & 0.0000 & 0.0069 & 0.6333 \\
FLRG→FLMB & 0.0102 & 0.0000 & 0.0070 & 0.7100 \\
FLMB→FLRG & 0.0411 & 0.0231 & 0.0086 & 0.0200* \\
FLRG→GRNB & 0.0413 & 0.0275 & 0.0070 & 0.0033** \\
GRNB→FLRG & 0.0335 & 0.0143 & 0.0086 & 0.0567. \\
FLRG→XGBU.SW & 0.0154 & 0.0000 & 0.0077 & 0.4300 \\
XGBU.SW→FLRG & 0.0135 & 0.0000 & 0.0083 & 0.5833 \\
GBNG.L→GRNB & 0.0378 & 0.0231 & 0.0073 & 0.0100* \\
GRNB→GBNG.L & 0.0268 & 0.0119 & 0.0071 & 0.0567. \\
GBNG.L→XGBU.SW & 0.0149 & 0.0000 & 0.0069 & 0.4667 \\
XGBU.SW→GBNG.L & 0.0135 & 0.0000 & 0.0071 & 0.3800 \\
GRNB→KLMH.F & 0.0569 & 0.0385 & 0.0089 & 0.0000 *** \\
KLMH.F→GRNB & 0.0279 & 0.0129 & 0.0066 & 0.0500. \\
GRNB→XCO2 & 0.0379 & 0.0207 & 0.0076 & 0.0100* \\
XCO2→GRNB & 0.0212 & 0.0073 & 0.0078 & 0.1800 \\
GRNB→XGBE.DE & 0.0187 & 0.0004 & 0.0084 & 0.3567 \\
XGBE.DE→GRNB & 0.0166 & 0.0020 & 0.0068 & 0.2667 \\
GRON.MI→GRNB & 0.0140 & 0.0000 & 0.0071 & 0.4033 \\
GRNB→GRON.MI & 0.0210 & 0.0052 & 0.0069 & 0.1867 \\
GRON.MI→XGBU.SW & 0.0161 & 0.0000 & 0.0072 & 0.3367 \\
XGBU.SW→GRON.MI & 0.0126 & 0.0000 & 0.0076 & 0.5933 \\
KLMH.F→XGBU.SW & 0.0202 & 0.0044 & 0.0072 & 0.2233 \\
XGBU.SW→KLMH.F & 0.0434 & 0.0254 & 0.0089 & 0.0133* \\
XCO2→XGBU.SW & 0.0141 & 0.0000 & 0.0066 & 0.4733 \\
XGBU.SW→XCO2 & 0.0143 & 0.0000 & 0.0081 & 0.5067 \\
XGBU.SW→XGBE.DE & 0.0183 & 0.0003 & 0.0078 & 0.3067 \\
XGBE.DE→XGBU.SW & 0.0212 & 0.0053 & 0.0073 & 0.2033 \\
\end{longtable}

\begin{figure}[H]
  \centering
    \begin{subfigure}{0.80\textwidth}
\centering
    \includegraphics[width=\linewidth]{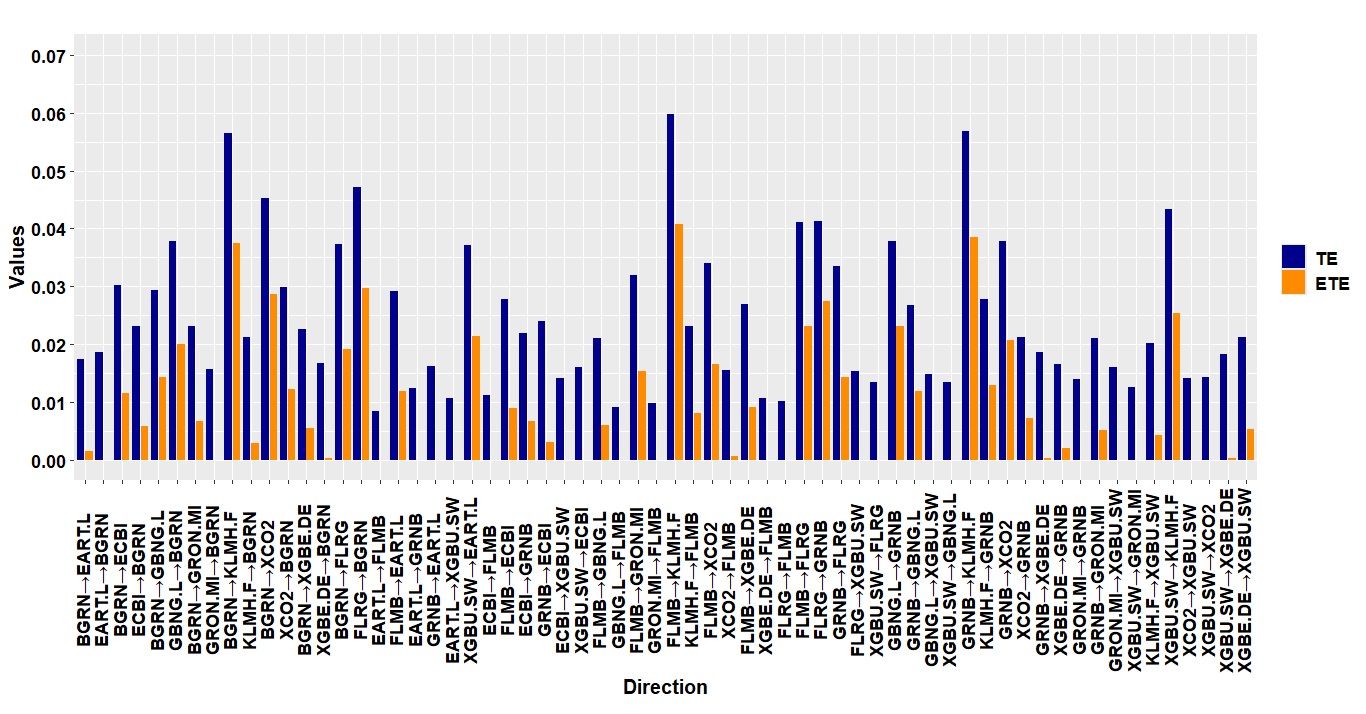}
         \caption{TE and ETE results among European and American ETFs.}
        \label{fig:Rplot03}
    \end{subfigure}
        \label{fig:european}
    \begin{subfigure}{0.80\textwidth}
        \centering
    \includegraphics[width=\linewidth]{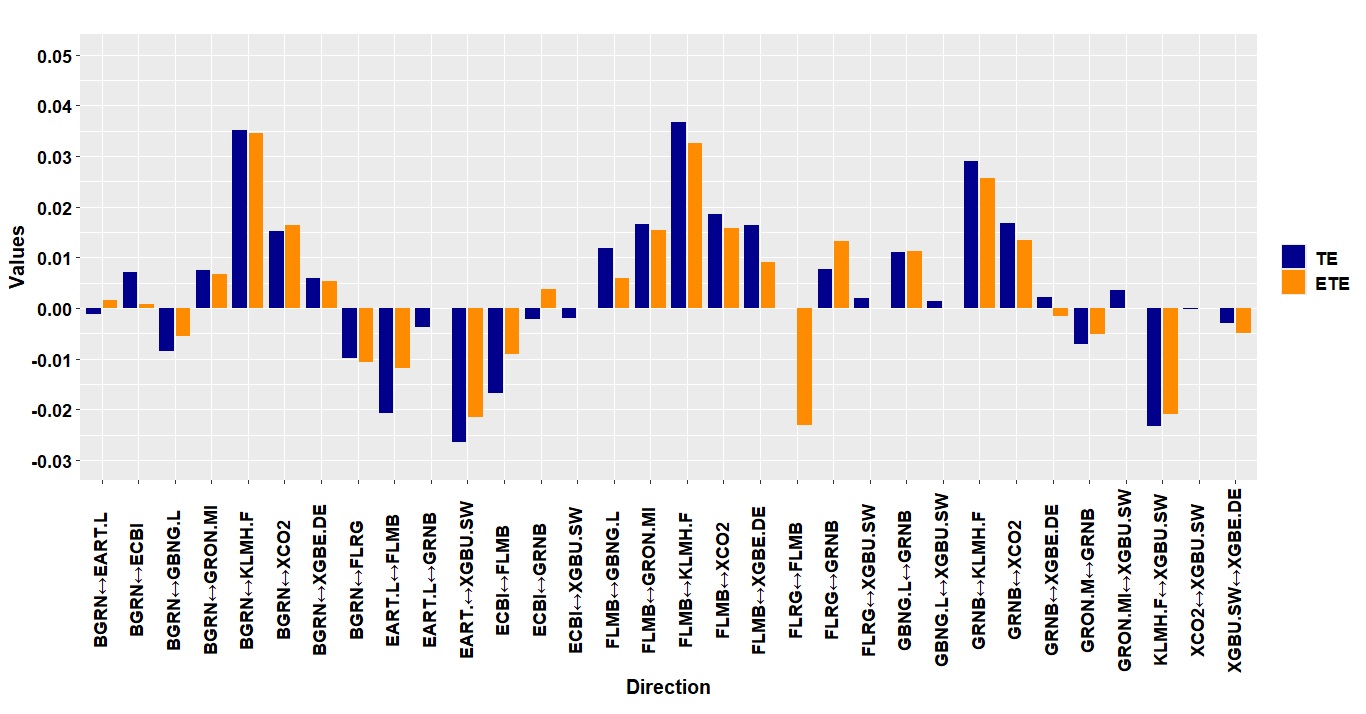} 
         \caption{TE and ETE difference among European and American ETFs.}
        \label{fig:europeanamerican10}
    \end{subfigure}
     \caption{TE and ETE results and difference among European and American ETFs.}
\end{figure}

The results of Table \ref{tab:EuropeanandAmerican} seen in Figure \ref{fig:Rplot03} indicate the asymmetry in the flow of information that exists between the flow of information between the USA and Europe. The differences shown in Figure \ref{fig:europeanamerican10} confirm what was observed in Figures \ref{fig:Rplot00} and \ref{fig:Rplot00american}, where FLMB ETF becomes a dominant EFT among the analyzed ETFs and ETF KLMH.F receives information from almost all other ETFs, but American EFTs like BGRN and XGBU.SW receives a lot of information from European ETFs, and the ETF GRNB contributes to the USA being the main issue among the locations studied.

\begin{figure}[H]
    \centering
    \includegraphics[width=0.9\linewidth]{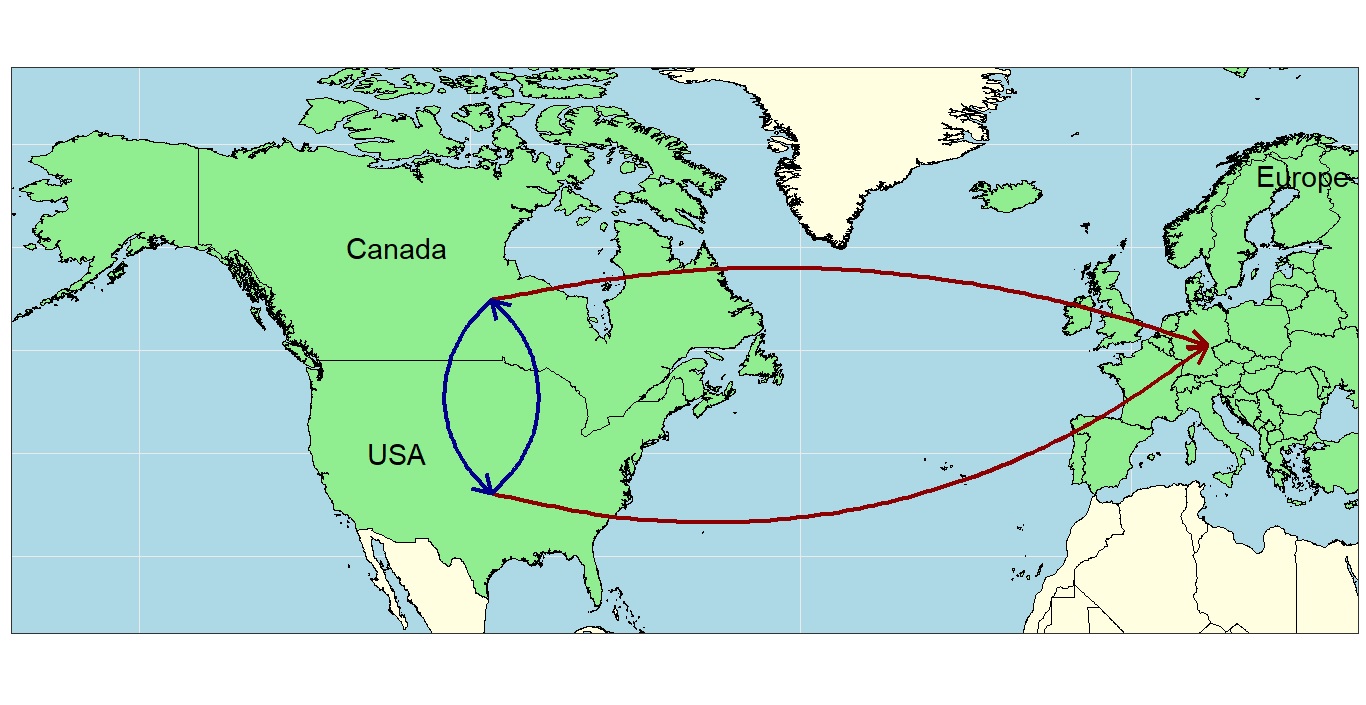}
    \caption{Flow of market information.}
    \label{fig:mapa}
\end{figure}
Based on the transfer of statistically significant information and confirmation of TE flow through the ETE described in the tables and figures above, we were able to map the flow of information in Figure \ref{fig:mapa} between the three markets analyzed in this study, where the USA remains dominant with its FLMB ETF that presents a high value of kurtosis (15.79349), sending information to Canada and Europe, and gaining influence within the territory itself.  Canada surprised and sent much statistically significant information to Europe, sending statistically significant information to the USA GRNB ETF. Europe has been open to receiving information from other countries, receiving a large flow of information from ETFs from the USA and Canada, and within its borders, the KLMH.F ETF receives information from other ETFs in its territory.

The study shows that the American market still has a strong influence on other markets, corroboration of Reboredo and Ugolini \cite{reboredo2020price} and He and Shang \cite{he2017comparison} studies that the American market has a strong influence on other markets, mainly through its Treasury ballbuster study, shows that the Canadian market can also influence other markets such as the USA and Europe and that decisions of countries on the other side of the Atlantic have influenced Europe.

\section{CONCLUSION}\label{sec:conclu}

%The advance of climate change increased the concerns of human beings in ways to mitigate the severe effects of these changes, for that created ways to encourage the realization of projects to mitigate these effects on the economy and life population, these Green Bonds have been playing its role to raise funds for these projects. For individuals to have access to invest in these projects, the well-known index funds with ETFs linked to Green Bonds were launched so investors can buy shares in these projects as if they were investing in stocks, For this reason, it is important to study understanding this market so that more people have access and seek ETFs Green Bonds.
% For this, three markets—American, European, and Canadian—were chosen to understand how information flows between them. 
The advance of climate change has increased human concerns about ways to mitigate its severe effects. To this end, it has created ways of encouraging the implementation of projects that reduce these effects on the economy and the population's lives. Green Bonds have played an important role in raising funds for these projects. For individuals to have access to invest in these projects, well-known index funds with Green Bond-linked ETFs were launched so that investors could buy shares in these projects as if they were investing in shares; therefore, it is important to study the understanding of this market so that more people have access and seek Green Bonds from ETFs.

This work investigated the non-linear theory of the directional information flow between the log-return series of 13 Green Bonds ETFs from three distinct markets: American, European, and Canadian. The results show that the information asymmetry is predominant between the USA, Canadian, and European markets, where the transfer of information does not have a defined balance, evidencing the unidimensionality among the vast majority of funds.

The Canadian HGGB ETF greatly influences European ETFs by sending significant information to the European 
KLMH.F, EART.L ETFs, and XCO2, having a unidimensional character, and the feedback of Europeans to it has no statistical significance. It showed its predominant influence on the European market. When information transfer from the American market to the Canadian market and vice versa is analyzed, the strong influence of the FLMB ETF on the HGGB ETF reveals the impact of the American market on the Canadian market.

The European market is perceived as a domain of FLRG and GRON.MI ETFs in information transfer, with great dominance in other funds such as KLMH.L and EART.L, there is also the predominance of one-dimensionality in which the transfer of information is balanced. Sending information between the European and American markets demonstrates the leadership of the US market in sending information between markets, where we can trace a path of information in the global markets that starts in the American market and sends information to the Canadian market, which after receiving it sends to Europe or that the American market can send directly to the European market. EFTs show that there is a way that information passes through, but this has an asymmetry in which there is a predominance of ETFs that send information. Others who send information are strongly affected, contributing to the construction of the predictable information path.

Thus, our work contributes to a better understanding of the dynamics of information flow between the three markets, which can help investors and managers mitigate market risks and contribute to the public debate on the role of countries and companies in creating sustainability projects. In future work, we can understand the dynamics between ETFs in Asian countries, such as China and India, and investigate the influence of ETF Kurtosis on the flow of information to better understand this relationship with TE and ETE.

\section*{ACKNOWLEDGMENTS}
%We acknowledge the support of the Brazilian agency CAPES.
We thank the Coordination for the Improvement of Higher Education Personnel - Brazil (CAPES) - Finance Code 001 for financial support for the development of this research.
\bibliography{bibliografia.bib}

\end{document}